\documentclass[letter,aps,prl,twocolumn,superscriptaddress,showpacs,preprintnumbers,amsmath,amssymb,nofootinbib]{revtex4-1}

\usepackage{graphicx} 
\usepackage{dcolumn}  

\usepackage{mathtools}

\usepackage{xspace}

\usepackage{dcolumn}

\input{babarsym.tex}

\graphicspath{{ps}}

\def\MDzero{\ensuremath{M_{\Dz}}\xspace}
\def\deltaMDstarDzero{\ensuremath{{\rm \Delta}M}\xspace}

\def\piplussoft {\ensuremath{\pip_{s}}\xspace}

\def\MsquaredKSpiRS{\ensuremath{M_{ \KS \pi^{-} }^{2}}\xspace}
\def\MsquaredKSpiWS{\ensuremath{M_{ \KS \pi^{+} }^{2}}\xspace}
\def\Msquaredpipi{\ensuremath{M_{ \pi^{+} \pi^{-} }^{2}}\xspace}

\def\Mbcprime{\ensuremath{M_{\mathrm{bc}}^{\prime}}\xspace}
\def\DeltaE{\ensuremath{\Delta E}\xspace}
\def\NNout{\ensuremath{\mathcal{C}_{N\kern-0.2emN_{\mathrm{out}}}}\xspace}
\def\NNoutmin{\ensuremath{\mathcal{C}^{\mathrm{min}}_{N\kern-0.2emN_{\mathrm{out}}}}\xspace}
\def\NNoutmax{\ensuremath{\mathcal{C}^{\mathrm{max}}_{N\kern-0.2emN_{\mathrm{out}}}}\xspace}
\def\NNoutprime{\ensuremath{\mathcal{C}_{N\kern-0.2emN_{\mathrm{out}}}^{\prime}}\xspace}

\def\Deltat{\ensuremath{\Delta t}\xspace}

\def\DstarplustoDzeropisoft {\ensuremath{ \Dstarp \to \Dz \piplussoft}\xspace}

\def\DzerotoKSpipi {\ensuremath{\Dz \to \KS \pip \pim}\xspace}

\def\DtoKSpipi {\ensuremath{\D \to \KS \pip \pim}\xspace}

\def\BtoDhzero {\ensuremath{\Bz \to \D^{(*)} h^{0}}\xspace}

\newcommand*{\sPlot}{\ensuremath{{}_s{\mathcal P}\mkern-2mu\mathit{lot}}\xspace}

\begin{document}

\title{ \quad\\[1.0cm] First evidence for {\boldmath{$\cos 2\beta>0$}} and resolution of the CKM Unitarity Triangle ambiguity by a time-dependent Dalitz plot analysis of {\boldmath$\Bz \to \D^{(*)} h^{0}$} with {\boldmath$\D \to \KS \pip \pim$} decays}

\newcommand{\TAGbbr}{${^\mathrm{A}}$}
\newcommand{\TAGbel}{${^\mathrm{B}}$}
\noaffiliation

\affiliation{Laboratoire d'Annecy-le-Vieux de Physique des Particules (LAPP), Universit\'e de Savoie, CNRS/IN2P3,  F-74941 Annecy-Le-Vieux, France}

\affiliation{Universitat de Barcelona, Facultat de Fisica, Departament ECM, E-08028 Barcelona, Spain }

\affiliation{INFN Sezione di Bari and Dipartimento di Fisica, Universit\`a di Bari, I-70126 Bari, Italy }

\affiliation{University of the Basque Country UPV/EHU, 48080 Bilbao, Spain }

\affiliation{Beihang University, Beijing 100191, China }

\affiliation{University of Bergen, Institute of Physics, N-5007 Bergen, Norway }

\affiliation{Lawrence Berkeley National Laboratory and University of California, Berkeley, California 94720, USA }

\affiliation{Ruhr Universit\"at Bochum, Institut f\"ur Experimentalphysik 1, D-44780 Bochum, Germany }

\affiliation{Institute of Particle Physics$^{\,a}$; University of British Columbia$^{b}$, Vancouver, British Columbia, Canada V6T 1Z1 }

\affiliation{Brookhaven National Laboratory, Upton, New York 11973, USA }

\affiliation{Budker Institute of Nuclear Physics SB RAS, Novosibirsk 630090, Russian Federation }

\affiliation{Novosibirsk State University, Novosibirsk 630090, Russian Federation }

\affiliation{Novosibirsk State Technical University, Novosibirsk 630092, Russian Federation }

\affiliation{University of California at Irvine, Irvine, California 92697, USA }

\affiliation{University of California at Riverside, Riverside, California 92521, USA }

\affiliation{University of California at Santa Cruz, Institute for Particle Physics, Santa Cruz, California 95064, USA }

\affiliation{California Institute of Technology, Pasadena, California 91125, USA }

\affiliation{Faculty of Mathematics and Physics, Charles University, 121 16 Prague, Czech Republic }

\affiliation{Chonnam National University, Kwangju 660-701, South Korea }

\affiliation{University of Cincinnati, Cincinnati, Ohio 45221, USA }

\affiliation{University of Colorado, Boulder, Colorado 80309, USA }

\affiliation{Deutsches Elektronen--Synchrotron, 22607 Hamburg, Germany }

\affiliation{Key Laboratory of Nuclear Physics and Ion-beam Application (MOE) and Institute of Modern Physics, Fudan University, Shanghai 200443, China }

\affiliation{INFN Sezione di Ferrara$^{a}$; Dipartimento di Fisica e Scienze della Terra, Universit\`a di Ferrara$^{b}$, I-44122 Ferrara, Italy }

\affiliation{INFN Laboratori Nazionali di Frascati, I-00044 Frascati, Italy }

\affiliation{INFN Sezione di Genova, I-16146 Genova, Italy }

\affiliation{Justus-Liebig-Universit\"at Gie\ss{}en, 35392 Gie\ss{}en, Germany }

\affiliation{SOKENDAI (The Graduate University for Advanced Studies), Hayama 240-0193, Japan }

\affiliation{Hanyang University, Seoul 133-791, South Korea }

\affiliation{University of Hawaii, Honolulu, Hawaii 96822, USA }

\affiliation{High Energy Accelerator Research Organization (KEK), Tsukuba 305-0801, Japan }

\affiliation{J-PARC Branch, KEK Theory Center, High Energy Accelerator Research Organization (KEK), Tsukuba 305-0801, Japan }

\affiliation{Humboldt-Universit\"at zu Berlin, Institut f\"ur Physik, D-12489 Berlin, Germany }

\affiliation{IKERBASQUE, Basque Foundation for Science, 48013 Bilbao, Spain }

\affiliation{Indian Institute of Science Education and Research Mohali, SAS Nagar, 140306, India }

\affiliation{Indian Institute of Technology Bhubaneswar, Satya Nagar 751007, India }

\affiliation{Indian Institute of Technology Guwahati, Assam 781039, India }

\affiliation{Indian Institute of Technology Hyderabad, Telangana 502285, India }

\affiliation{Indian Institute of Technology Madras, Chennai 600036, India }

\affiliation{Indiana University, Bloomington, Indiana 47408, USA }

\affiliation{Institute of High Energy Physics, Chinese Academy of Sciences, Beijing 100049, China }

\affiliation{Institute of High Energy Physics, Vienna 1050, Austria }

\affiliation{University of Iowa, Iowa City, Iowa 52242, USA }

\affiliation{Iowa State University, Ames, Iowa 50011, USA }

\affiliation{Advanced Science Research Center, Japan Atomic Energy Agency, Naka 319-1195}

\affiliation{J. Stefan Institute, 1000 Ljubljana, Slovenia }

\affiliation{Johns Hopkins University, Baltimore, Maryland 21218, USA }

\affiliation{Kanagawa University, Yokohama 221-8686, Japan }

\affiliation{Institut f\"ur Experimentelle Teilchenphysik, Karlsruher Institut f\"ur Technologie, 76131 Karlsruhe, Germany }

\affiliation{Kennesaw State University, Kennesaw, Georgia 30144, USA }

\affiliation{King Abdulaziz City for Science and Technology, Riyadh 11442, Kingdom of Saudi Arabia }

\affiliation{Department of Physics, Faculty of Science, King Abdulaziz University, Jeddah 21589, Kingdom of Saudi Arabia }

\affiliation{Korea Institute of Science and Technology Information, Daejeon 305-806, South Korea }

\affiliation{Korea University, Seoul 136-713, South Korea }

\affiliation{Kyungpook National University, Daegu 702-701, South Korea }

\affiliation{Laboratoire de l'Acc\'el\'erateur Lin\'eaire, IN2P3/CNRS et Universit\'e Paris-Sud 11, Centre Scientifique d'Orsay, F-91898 Orsay Cedex, France }

\affiliation{\'Ecole Polytechnique F\'ed\'erale de Lausanne (EPFL), Lausanne 1015, Switzerland }

\affiliation{Lawrence Livermore National Laboratory, Livermore, California 94550, USA }

\affiliation{P.N. Lebedev Physical Institute of the Russian Academy of Sciences, Moscow 119991, Russian Federation }

\affiliation{Laboratoire Leprince-Ringuet, Ecole Polytechnique, CNRS/IN2P3, F-91128 Palaiseau, France }

\affiliation{University of Liverpool, Liverpool L69 7ZE, United Kingdom }

\affiliation{Faculty of Mathematics and Physics, University of Ljubljana, 1000 Ljubljana, Slovenia }

\affiliation{Queen Mary, University of London, London, E1 4NS, United Kingdom }

\affiliation{University of London, Royal Holloway and Bedford New College, Egham, Surrey TW20 0EX, United Kingdom }

\affiliation{University of Louisville, Louisville, Kentucky 40292, USA }

\affiliation{Ludwig Maximilians University, 80539 Munich, Germany }

\affiliation{Luther College, Decorah, Iowa 52101, USA }

\affiliation{Johannes Gutenberg-Universit\"at Mainz, Institut f\"ur Kernphysik, D-55099 Mainz, Germany }

\affiliation{University of Malaya, 50603 Kuala Lumpur, Malaysia }

\affiliation{University of Manchester, Manchester M13 9PL, United Kingdom }

\affiliation{University of Maribor, 2000 Maribor, Slovenia }

\affiliation{University of Maryland, College Park, Maryland 20742, USA }

\affiliation{Massachusetts Institute of Technology, Laboratory for Nuclear Science, Cambridge, Massachusetts 02139, USA }

\affiliation{Max-Planck-Institut f\"ur Physik, 80805 M\"unchen, Germany }

\affiliation{Institute of Particle Physics$^{\,a}$; McGill University$^{b}$, Montr\'eal, Qu\'ebec, Canada H3A 2T8 }

\affiliation{School of Physics, University of Melbourne, Victoria 3010, Australia }

\affiliation{INFN Sezione di Milano$^{a}$; Dipartimento di Fisica, Universit\`a di Milano$^{b}$, I-20133 Milano, Italy }

\affiliation{University of Mississippi, University, Mississippi 38677, USA }

\affiliation{University of Miyazaki, Miyazaki 889-2192, Japan }

\affiliation{Universit\'e de Montr\'eal, Physique des Particules, Montr\'eal, Qu\'ebec, Canada H3C 3J7  }

\affiliation{Moscow Physical Engineering Institute, Moscow 115409, Russian Federation }

\affiliation{Moscow Institute of Physics and Technology, Moscow Region 141700, Russian Federation }

\affiliation{Graduate School of Science, Nagoya University, Nagoya 464-8602, Japan }

\affiliation{INFN Sezione di Napoli$^{a}$ and Dipartimento di Scienze Fisiche, Universit\`a di Napoli Federico II$^{b}$, I-80126 Napoli, Italy }

\affiliation{Nara Women's University, Nara 630-8506, Japan }

\affiliation{National Central University, Chung-li 32054, Taiwan }

\affiliation{National United University, Miao Li 36003, Taiwan }

\affiliation{NIKHEF, National Institute for Nuclear Physics and High Energy Physics, NL-1009 DB Amsterdam, The Netherlands }

\affiliation{Department of Physics, National Taiwan University, Taipei 10617, Taiwan }

\affiliation{H. Niewodniczanski Institute of Nuclear Physics, Krakow 31-342, Poland }

\affiliation{Nippon Dental University, Niigata 951-8580, Japan }

\affiliation{Niigata University, Niigata 950-2181, Japan }

\affiliation{University of Notre Dame, Notre Dame, Indiana 46556, USA }

\affiliation{Ohio State University, Columbus, Ohio 43210, USA }

\affiliation{Novosibirsk State University, Novosibirsk 630090, Russian Federation }

\affiliation{Osaka City University, Osaka 558-8585, Japan }

\affiliation{Pacific Northwest National Laboratory, Richland, Washington 99352, USA }

\affiliation{INFN Sezione di Padova$^{a}$; Dipartimento di Fisica, Universit\`a di Padova$^{b}$, I-35131 Padova, Italy }

\affiliation{Panjab University, Chandigarh 160014, India }

\affiliation{Laboratoire de Physique Nucl\'eaire et de Hautes Energies, IN2P3/CNRS, Universit\'e Pierre et Marie Curie-Paris6, Universit\'e Denis Diderot-Paris7, F-75252 Paris, France }

\affiliation{Peking University, Beijing 100871, China }

\affiliation{INFN Sezione di Perugia$^{a}$; Dipartimento di Fisica, Universit\`a di Perugia$^{b}$, I-06123 Perugia, Italy }

\affiliation{INFN Sezione di Pisa$^{a}$; Dipartimento di Fisica, Universit\`a di Pisa$^{b}$; Scuola Normale Superiore di Pisa$^{c}$, I-56127 Pisa, Italy }

\affiliation{University of Pittsburgh, Pittsburgh, Pennsylvania 15260, USA }

\affiliation{Princeton University, Princeton, New Jersey 08544, USA }

\affiliation{Theoretical Research Division, Nishina Center, RIKEN, Saitama 351-0198, Japan }

\affiliation{INFN Sezione di Roma$^{a}$; Dipartimento di Fisica, Universit\`a di Roma La Sapienza$^{b}$, I-00185 Roma, Italy }

\affiliation{Universit\"at Rostock, D-18051 Rostock, Germany }

\affiliation{Rutherford Appleton Laboratory, Chilton, Didcot, Oxon, OX11 0QX, United Kingdom }

\affiliation{CEA, Irfu, SPP, Centre de Saclay, F-91191 Gif-sur-Yvette, France }

\affiliation{University of Science and Technology of China, Hefei 230026, China }

\affiliation{Showa Pharmaceutical University, Tokyo 194-8543, Japan }

\affiliation{Soongsil University, Seoul 156-743, South Korea }

\affiliation{SLAC National Accelerator Laboratory, Stanford, California 94309 USA }

\affiliation{University of South Carolina, Columbia, South Carolina 29208, USA }

\affiliation{University of South Carolina, Columbia, South Carolina 29208, USA }

\affiliation{Southern Methodist University, Dallas, Texas 75275, USA }

\affiliation{St. Francis Xavier University, Antigonish, Nova Scotia, Canada B2G 2W5 }

\affiliation{Stanford University, Stanford, California 94305, USA }

\affiliation{State University of New York, Albany, New York 12222, USA }

\affiliation{Sungkyunkwan University, Suwon 440-746, South Korea }

\affiliation{School of Physics, University of Sydney, New South Wales 2006, Australia }

\affiliation{Department of Physics, Faculty of Science, University of Tabuk, Tabuk 71451, Kingdom of Saudi Arabia }

\affiliation{Tata Institute of Fundamental Research, Mumbai 400005, India }

\affiliation{Department of Physics, Technische Universit\"at M\"unchen, 85748 Garching, Germany }

\affiliation{Excellence Cluster Universe, Technische Universit\"at M\"unchen, 85748 Garching, Germany }

\affiliation{Tel Aviv University, School of Physics and Astronomy, Tel Aviv, 69978, Israel }

\affiliation{University of Tennessee, Knoxville, Tennessee 37996, USA }

\affiliation{University of Texas at Austin, Austin, Texas 78712, USA }

\affiliation{University of Texas at Dallas, Richardson, Texas 75083, USA }

\affiliation{Toho University, Funabashi 274-8510, Japan }

\affiliation{Department of Physics, Tohoku University, Sendai 980-8578, Japan }

\affiliation{Earthquake Research Institute, University of Tokyo, Tokyo 113-0032, Japan }

\affiliation{Department of Physics, University of Tokyo, Tokyo 113-0033, Japan }

\affiliation{Tokyo Institute of Technology, Tokyo 152-8550, Japan }

\affiliation{Tokyo Metropolitan University, Tokyo 192-0397, Japan }

\affiliation{INFN Sezione di Torino$^{a}$; Dipartimento di Fisica, Universit\`a di Torino$^{b}$, I-10125 Torino, Italy }

\affiliation{INFN Sezione di Trieste and Dipartimento di Fisica, Universit\`a di Trieste, I-34127 Trieste, Italy }

\affiliation{IFIC, Universitat de Valencia-CSIC, E-46071 Valencia, Spain }

\affiliation{Institute of Particle Physics$^{\,a}$; University of Victoria$^{b}$, Victoria, British Columbia, Canada V8W 3P6 }

\affiliation{Virginia Polytechnic Institute and State University, Blacksburg, Virginia 24061, USA }

\affiliation{Department of Physics, University of Warwick, Coventry CV4 7AL, United Kingdom }

\affiliation{Wayne State University, Detroit, Michigan 48202, USA }

\affiliation{University of Wisconsin, Madison, Wisconsin 53706, USA }

\affiliation{Yamagata University, Yamagata 990-8560, Japan }

\affiliation{Yonsei University, Seoul 120-749, South Korea }

\author{I.~Adachi\TAGbel}\affiliation{High Energy Accelerator Research Organization (KEK), Tsukuba 305-0801, Japan }\affiliation{SOKENDAI (The Graduate University for Advanced Studies), Hayama 240-0193, Japan } 
\author{T.~Adye\TAGbbr}\affiliation{Rutherford Appleton Laboratory, Chilton, Didcot, Oxon, OX11 0QX, United Kingdom }
\author{H.~Ahmed\TAGbbr}\affiliation{St. Francis Xavier University, Antigonish, Nova Scotia, Canada B2G 2W5 }
\author{J.~K.~Ahn\TAGbel}\affiliation{Korea University, Seoul 136-713, South Korea } 
\author{H.~Aihara\TAGbel}\affiliation{Department of Physics, University of Tokyo, Tokyo 113-0033, Japan } 
\author{S.~Akar\TAGbbr}\affiliation{Laboratoire de Physique Nucl\'eaire et de Hautes Energies, IN2P3/CNRS, Universit\'e Pierre et Marie Curie-Paris6, Universit\'e Denis Diderot-Paris7, F-75252 Paris, France }
\author{M.~S.~Alam\TAGbbr}\affiliation{State University of New York, Albany, New York 12222, USA }
\author{J.~Albert\TAGbbr$^{b}$}\affiliation{Institute of Particle Physics$^{\,a}$; University of Victoria$^{b}$, Victoria, British Columbia, Canada V8W 3P6 }
\author{F.~Anulli\TAGbbr$^{a}$}\affiliation{INFN Sezione di Roma$^{a}$; Dipartimento di Fisica, Universit\`a di Roma La Sapienza$^{b}$, I-00185 Roma, Italy }
\author{N.~Arnaud\TAGbbr}\affiliation{Laboratoire de l'Acc\'el\'erateur Lin\'eaire, IN2P3/CNRS et Universit\'e Paris-Sud 11, Centre Scientifique d'Orsay, F-91898 Orsay Cedex, France }
% ...

\collaboration{The {\TAGbbr}\babar\ and {\TAGbel}Belle Collaborations}

\begin{abstract}
We present first evidence that the cosine of the \CP-violating weak phase $2\beta$ is positive, and hence exclude trigonometric multifold solutions of the CKM Unitarity Triangle
using a time-dependent Dalitz plot analysis of \BtoDhzero with \DtoKSpipi decays, where $h^{0} \in \{\pi^{0}, \eta, \omega \}$ denotes a light unflavored and neutral hadron.
The measurement is performed combining the final data sets of the \babar\ and Belle experiments collected at the $\Upsilon(4S)$ resonance at the asymmetric-energy \B factories PEP-II at SLAC and KEKB at KEK, respectively.
The data samples contain $( 471 \pm 3 )\times 10^6\, \B\Bbar$ pairs recorded by the \babar\ detector and $( 772 \pm 11 )\times 10^6\, \B\Bbar$ pairs recorded by the Belle detector.
The results of the measurement are $\sin{2\beta} = 0.80 \pm 0.14  \,(\rm{stat.}) \pm 0.06 \,(\rm{syst.}) \pm 0.03 \,(\rm{model})$ and $\cos{2\beta} = 0.91 \pm 0.22  \,(\rm{stat.}) \pm 0.09 \,(\rm{syst.}) \pm 0.07 \,(\rm{model})$.
The result for the direct measurement of the angle $\beta$ of the CKM Unitarity Triangle is $\beta = \left( 22.5 \pm 4.4  \,(\rm{stat.}) \pm 1.2 \,(\rm{syst.}) \pm 0.6 \,(\rm{model}) \right)^{\circ}$.
The quoted model uncertainties are due to the composition of the \DzerotoKSpipi decay amplitude model, which is newly established by performing a Dalitz plot amplitude analysis using a high-statistics $e^{+}e^{-} \to c\bar{c}$ data sample.
\CP violation is observed in \BtoDhzero decays at the level of $5.1$ standard deviations. The significance for $\cos{2\beta}>0$ is $3.7$ standard deviations.
The trigonometric multifold solution $\pi/2 - \beta = (68.1 \pm 0.7)^{\circ}$ is excluded at the level of $7.3$ standard deviations.
The measurement resolves an ambiguity in the determination of the apex of the CKM Unitarity Triangle.
\end{abstract}

\pacs{11.30.Er, 12.15.Hh, 13.25.Hw}

\maketitle

\tighten

{\renewcommand{\thefootnote}{\fnsymbol{footnote}}}
\setcounter{footnote}{0}

In the standard model (SM) of electroweak interactions, the only source of \CP violation is the irreducible complex phase in the three-family Cabibbo-Kobayashi-Maskawa (CKM) quark-mixing matrix~\cite{CabibboKobayashiMaskawa}.
The \babar\ and Belle experiments discovered \CP violation in the \B meson system~\cite{CPV_observation_BaBar,CPV_observation_Belle,directCPV_BaBar,directCPV_Belle}.
In particular, by time-dependent \CP violation measurements of the ``gold plated'' decay mode\footnote{In this Letter the inclusion of charge-conjugated decay modes is implied unless otherwise stated.} $\Bz \to J/\psi K_{S}^{0}$ and other decays mediated by $\bar{b} \to \bar{c}c\bar{s}$ transitions~\cite{BaBar_btoccs,Belle_btoccs},
\babar\ and Belle precisely determined the parameter $\sin{2\beta} \equiv \sin{2\phi_1}$,\footnote{\babar\ uses the notation $\beta$ and Belle uses $\phi_1$; hereinafter $\beta$ is used.}
where the angle $\beta$ of the CKM Unitarity Triangle is defined as $\arg\left[-V^{}_{cd} V^{*}_{cb} / V^{}_{td} V^{*}_{tb} \right]$ and $V_{ij}$ denotes a CKM matrix element.
Inferring the \CP-violating weak phase $2\beta$ from these measurements of $\sin{2\beta}$ leads to the trigonometric two-fold ambiguity, $2\beta$ and $\pi - 2\beta$ (a four-fold ambiguity in $\beta$), and therefore to an ambiguity on the CKM Unitarity Triangle.
This ambiguity can be resolved by also measuring $\cos{2\beta}$, which is experimentally accessible in \B meson decay modes involving multibody final states
such as $\Bz \to J/\psi \KS \piz$~\cite{BaBar_BToJPsiK0Spi0_2005,Belle_BToJPsiK0Spi0_2005}, $\Bz \to \Dstarp \Dstarm \KS$~\cite{BaBar_BToDstarDstarK0S_2006,Belle_BToDstarDstarK0S_2007},
$\Bz \to \KS \Kp \Km$~\cite{BaBar_BToKSKK_2012,Belle_BToKSKK_2010}, $\Bz \to \KS \pip \pim$~\cite{BaBar_BToKSpipi_2009,Belle_BToKSpipi_2009},
and $\Bz \to \D^{(*)} h^{0}$ with $\D \to \KS \pip \pim$ decays (abbreviated as $\Bz \to \left[ \KS \pip \pim \right]^{(*)}_{D} h^{0}$)~\cite{Belle_D0h0_2006,BaBar_D0h0_2007_threebody,Belle_D0h0_2016}.
However, no previous single measurement has been sufficiently sensitive to establish the sign of $\cos{2\beta}$, to resolve the ambiguity without further assumptions.

The decays \BtoDhzero, with \DtoKSpipi and $h^{0} \in \{\pi^{0}, \eta, \omega \}$ denoting a light neutral hadron, provide an elegant way to access $\cos{2\beta}$~\cite{BondarGershonKrokovny2005}.
The \BtoDhzero decay is predominantly mediated by CKM-favored $\bar{b} \to \bar{c}u\bar{d}$ tree amplitudes.
Additional contributions from CKM-disfavored $\bar{b} \to \bar{u}c\bar{d}$ tree amplitudes that carry different weak phases 
are suppressed by $\lvert V^{}_{ub} V^{*}_{cd} / V^{}_{cb} V^{*}_{ud} \rvert \approx 0.02$ relative to the leading amplitudes and can be neglected at the experimental sensitivity of the presented measurement.
The \DtoKSpipi decay exhibits complex interference structures that receive resonant and nonresonant contributions to the three-body final state
from a rich variety of intermediate \CP eigenstates and quasi-flavor-specific decays.
Knowledge of the variations on the relative strong phase as a function of the three-body Dalitz plot phase space enables measurements of both $\sin{2\beta}$ and $\cos{2\beta}$
from the time evolution of the $\Bz \to \left[ \KS \pip \pim \right]^{(*)}_{D} h^{0}$ multibody final state.

Assuming no \CP violation in \Bz-\Bzb mixing and no direct \CP violation,
the rate of the $\Bz \to \left[ \KS \pip \pim \right]^{(*)}_{D} h^{0}$ decays is proportional to
\begin{align}
  & \frac{ e^{ \frac{-\lvert \Delta t \rvert}{\tau_{\Bz} } } }{2} \Big\{ \left[ \lvert \mathcal{A}_{\Dzb} \rvert^{2} + \lvert \mathcal{A}_{\Dz} \rvert^{2} \right] \nonumber \\
  & \quad - q \left( \lvert \mathcal{A}_{\Dzb} \rvert^{2} - \lvert \mathcal{A}_{\Dz} \rvert^{2} \right) \cos(\Delta m_{d} \Delta t) \nonumber \\
  & \quad + 2 q \eta_{h^{0}} \left( -1 \right)^{L} \mathrm{Im} \left( e^{-2 i \beta} \mathcal{A}_{\Dz} \mathcal{A}_{\Dzb}^{*} \right) \sin(\Delta m_{d} \Delta t) \Big\} \mathrm{,} \label{equation:decay_rate}
\end{align}
where \Deltat denotes the proper-time interval between the decays of the two \B mesons produced in the $e^+ e^- \to \Upsilon\left(4S\right) \to \Bz\Bzb$ event,
and $q = +1$ ($-1$) represents the $b$-flavor content when the accompanying \B meson is tagged as a \Bz(\Bzb).
The parameters $\tau_{\Bz}$ and $\Delta m_{d}$ are the neutral \B meson lifetime and the \Bz-\Bzb oscillation frequency, respectively.
The symbols $\mathcal{A}_{\Dz} \equiv \mathcal{A}( \MsquaredKSpiRS, \MsquaredKSpiWS )$ and $\mathcal{A}_{\Dzb} \equiv \mathcal{A}( \MsquaredKSpiWS, \MsquaredKSpiRS )$ denote the \Dz and \Dzb decay amplitudes
as functions of the Lorentz-invariant Dalitz plot variables $\MsquaredKSpiRS \equiv (p_{\KS} + p_{\pi^{-}})^{2}$ and $\MsquaredKSpiWS \equiv (p_{\KS} + p_{\pi^{+}})^{2}$,
where the symbol $p_i$ represents the four-momentum of a final state particle $i$.
The factor $\eta_{h^{0}}$ is the \CP eigenvalue of $h^{0}$.
The quantity $L$ is the orbital angular momentum of the $\D h^{0}$ or $\Dstar h^{0}$ system.
The last term in Eq.~(\ref{equation:decay_rate}) can be rewritten as
\begin{align}
  \mathrm{Im} \left( e^{-2 i \beta} \mathcal{A}_{\Dz} \mathcal{A}_{\Dzb}^{*} \right) = & \ \mathrm{Im} \left( \mathcal{A}_{\Dz} \mathcal{A}_{\Dzb}^{*} \right) \cos{2\beta} \nonumber \\
  & - \mathrm{Re} \left( \mathcal{A}_{\Dz} \mathcal{A}_{\Dzb}^{*} \right) \sin{2\beta} \mathrm{,} \label{equation_sine_cosine}
\end{align}
which allows $\sin{2\beta}$ and $\cos{2\beta}$ to be treated as independent parameters.

Measurements of $\sin{2\beta}$ and $\cos{2\beta}$ in \BtoDhzero with \DtoKSpipi decays are experimentally challenging.
The branching fractions of the \B and \D meson decays are low ($\mathcal{O}(10^{-4})$ and $\mathcal{O}(10^{-2})$, respectively),
and the neutral particles in the final state lead to large backgrounds and low reconstruction efficiencies.
In addition, a detailed Dalitz plot amplitude model or other experimental knowledge of the relative strong phase in the three-body \D meson decay is required.
Previous measurements of these decays performed separately by \babar\ and Belle were not sufficiently sensitive to establish \CP violation~\cite{Belle_D0h0_2006,BaBar_D0h0_2007_threebody,Belle_D0h0_2016},
obtaining results far outside of the physical region of the parameter space~\cite{Belle_D0h0_2006}, and using different Dalitz plot amplitude models~\cite{Belle_D0h0_2006,BaBar_D0h0_2007_threebody},
which complicates the combination of individual results.

In this Letter, we present measurements of $\sin{2\beta}$ and $\cos{2\beta}$ from a time-dependent Dalitz plot analysis of \BtoDhzero with \DtoKSpipi decays
that combines the final data samples collected by the \babar\ and Belle experiments, totaling $1.1\,\mathrm{ab}^{-1}$ collected at the $\Upsilon\left(4S\right)$ resonance.
The combined approach enables unique experimental sensitivity to $\cos{2\beta}$ by increasing the available data sample and by applying common assumptions and the same Dalitz plot amplitude model simultaneously to the data collected by both experiments.
As part of the analysis, an improved \DtoKSpipi Dalitz plot amplitude model is obtained from high-statistics $e^{+}e^{-} \to c\bar{c}$ data.
This allows the propagation of the model uncertainties to the results on $\sin{2\beta}$ and $\cos{2\beta}$ obtained in \BtoDhzero with \DzerotoKSpipi decays in a straightforward way.
In the following, the extraction of the \DzerotoKSpipi Dalitz plot amplitude model parameters from Belle $e^{+}e^{-} \to c\bar{c}$ data is summarized.
Thereafter, the time-dependent Dalitz plot analysis of the \B meson decay combining \babar\ and Belle, data is described.
A more detailed description of the analysis is provided in Ref.~\cite{accompanyingPRDpaper}.

To measure the \DzerotoKSpipi decay amplitudes, we use a data sample of $924\,\mathrm{fb}^{-1}$ recorded at or near the $\Upsilon(4S)$ and $\Upsilon(5S)$ resonances with the Belle detector~\cite{BelleDetector} at the asymmetric-energy $e^+ e^-$ collider KEKB~\cite{KEKB}.
This gives a large sample of \D mesons enabling precise measurement of the decay amplitudes, so in effect nothing would be gained by the inclusion of the equivalent \babar\ data.
The decays \DstarplustoDzeropisoft with \DzerotoKSpipi and $\KS \to \pip \pim$ are reconstructed, and the flavor of the neutral \D meson is identified as \Dz (\Dzb) by the positive (negative) charge of the slow pion \piplussoft emitted from the \Dstarp decay.
Charged pion candidates are formed from reconstructed tracks, and the selection requirements described in Refs.~\cite{BELLEKshortselection,Roehrken2015} are applied to \KS candidates.
To reject background originating from \B meson decays, a requirement of $p^{*}(\Dstarp) > 2.5\, (3.1)\gevc$ for candidates reconstructed from $\Upsilon(4S)$ ($\Upsilon(5S)$) data is applied,
where $p^{*}$ denotes the momentum evaluated in the $e^+ e^-$ center-of-mass (c.m.) frame.
Events are selected by the \Dz candidate mass \MDzero and the $D^{*+} - D^{0}$ mass difference \deltaMDstarDzero,
and a yield of $1\,217\,300 \pm 2\,000$ signal decays is obtained by a two-dimensional unbinned maximum-likelihood fit to the \MDzero and \deltaMDstarDzero distributions~\cite{accompanyingPRDpaper}.

Similar to previous \Dz-\Dzb oscillation analyses and measurements of the Unitarity Triangle angle $\gamma$~\cite{phi_three} by \babar, Belle and LHCb~\cite{BABAR2008,BABAR2010,Peng2014,LHCb2014},
the \DzerotoKSpipi decay amplitude is parameterized as:
\begin{widetext}
\begin{equation}
\mathcal{A}( \MsquaredKSpiRS, \MsquaredKSpiWS ) = \kern-2em \sum\limits_{r \neq (K\pi/\pi\pi)_{L=0} }^{} \kern-2em a_{r} e^{i \phi_{r}} \mathcal{A}_{r} ( \MsquaredKSpiRS, \MsquaredKSpiWS ) + F_{1} (\Msquaredpipi) + \mathcal{A}_{{K\pi}_{L=0}} (\MsquaredKSpiRS) + \mathcal{A}_{{K\pi}_{L=0}} (\MsquaredKSpiWS) \mathrm{.} \label{eqn:Dalitz_amplitude_model}  
\end{equation}
\end{widetext}
The symbols $a_{r}$ and $\phi_{r}$ represent the magnitude and phase of the $r^{\mathrm{th}}$ intermediate quasi-two-body amplitude $\mathcal{A}_{r}$ contributing to the $P$- and $D$-waves.
These amplitudes are parameterized using an isobar ansatz~\cite{ReviewDalitzPlotAnalysis} by relativistic Breit-Wigner (BW) propagators with mass-dependent widths,
Blatt-Weisskopf penetration factors~\cite{BlattWeisskopfFormFactors}, and Zemach tensors for the angular distributions~\cite{ZemachTensors}.
The following intermediate two-body resonances are included:
the Cabibbo-favored
$K^{*}(892)^{-} \pip$,
$K^{*}_{2}(1430)^{-} \pip$,
$K^{*}(1680)^{-} \pip$,
$K^{*}(1410)^{-} \pip$ channels;
the doubly Cabibbo-suppressed
$K^{*}(892)^{+} \pim$,
$K^{*}_{2}(1430)^{+} \pim$,
$K^{*}(1410)^{+} \pim$ modes;
and the \CP eigenstates
$\KS \rho(770)^{0}$,
$\KS \omega(782)$,
$\KS f_{2}(1270)$,
and
$\KS \rho(1450)^{0}$.
The symbol $F_{1}$ denotes the amplitude for the $\pi\pi$ $S$-wave using the $K$-matrix formalism in the $P$-vector approximation with 4 physical poles~\cite{Kmatrix1995,Kmatrix2003}.
The symbol $\mathcal{A}_{{K\pi}_{L=0}}$ represents the amplitude for the $K\pi$ $S$-wave using the LASS parametrization~\cite{LASS}, which combines a BW for the $K^{*}_{0}(1430)^\pm$ with a coherent nonresonant contribution governed by an effective range and a phase shift.

The \DzerotoKSpipi decay amplitude model parameters are determined by an unbinned maximum-likelihood Dalitz fit performed for events in the signal region of the flavor-tagged \Dz sample.
The probability density function (p.d.f.) for the signal is constructed from Eq.~(\ref{eqn:Dalitz_amplitude_model}) with a correction to account for reconstruction efficiency variations over
the Dalitz plot phase space due to experimental acceptance effects~\cite{MCgeneratorsandsimulations}, and an additional term to account for wrong flavor identifications of \D mesons.
In addition, the likelihood function contains a p.d.f.\  for the background that is constructed from the distributions taken from the \MDzero and \deltaMDstarDzero data sidebands.
The $a_{r}$ and $\phi_{r}$ parameters for each resonance are floated in the fit and measured relative to the $\KS \rho(770)^{0}$ amplitude,
which is fixed to $a_{\KS \rho(770)^{0}}=1$ and $\phi_{\KS \rho(770)^{0}}=0^\circ$.
The masses and widths of the resonances are fixed to the world averages~\cite{PDG2016} except for those of the $ K^{*}(892)$ and $ K^{*}_{0}(1430)$, which are floated to improve the fit quality.
The LASS parameters and several parameters in the $K$-matrix are floated in the fit.

\begin{figure}[htb]
\includegraphics[width=0.47\textwidth]{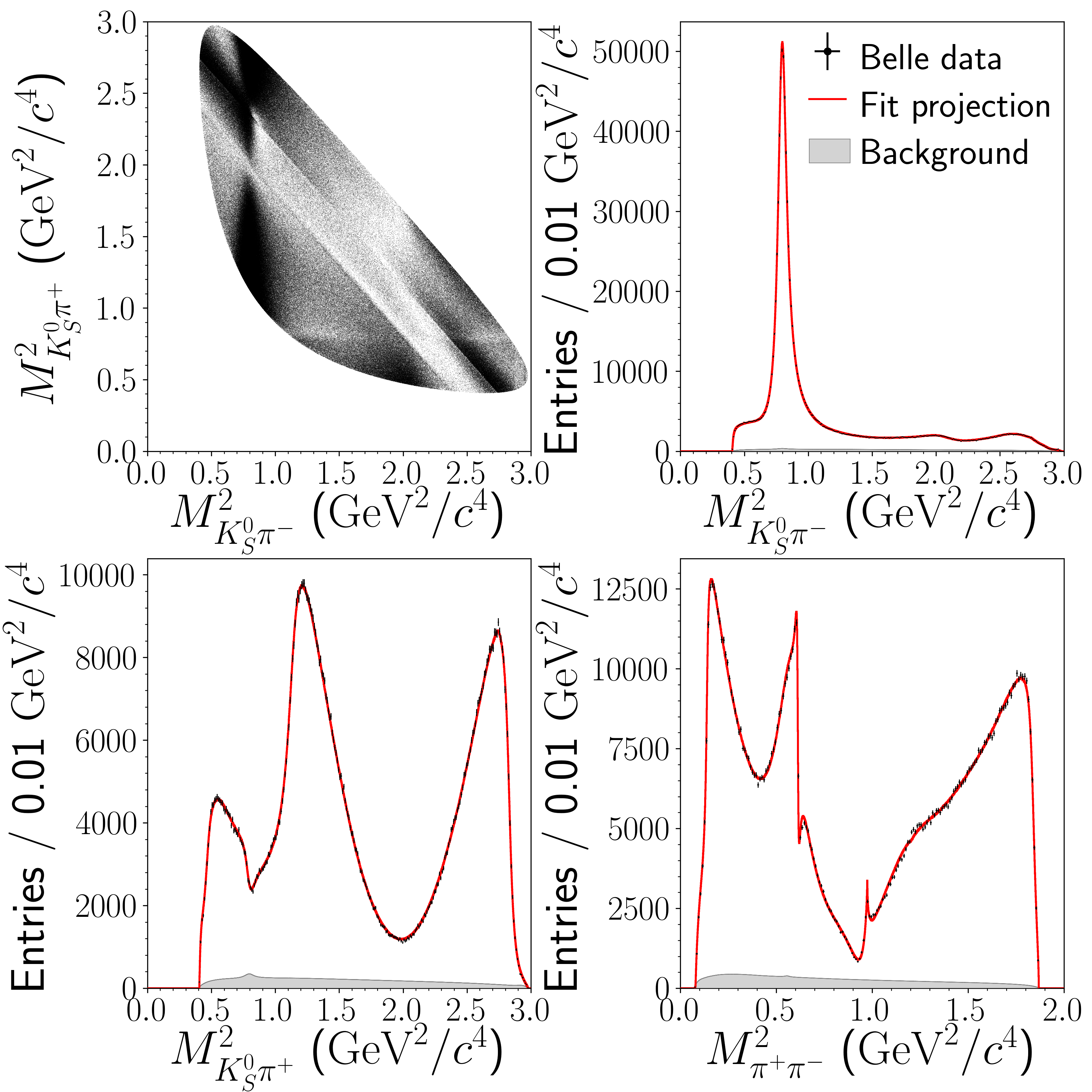}
\caption{
(color online). The Dalitz plot data distributions (points with error bars) for \DzerotoKSpipi from \DstarplustoDzeropisoft decays reconstructed from Belle ${e^{+}e^{-} \to c\bar{c}}$ data,
and projections of the Dalitz fit. The red solid lines show the projections of the total fit function including background,
and the grey regions show projections of the background.
}
\label{figure_Dalitz_fit}
\end{figure}

The results of the Dalitz fit are summarized in Table~{III} of Ref.~\cite{accompanyingPRDpaper}. The data distributions and projections of the fit are shown in Fig.~\ref{figure_Dalitz_fit}.
By a two-dimensional $\chi^{2}$ test, a reduced $\chi^{2}$ of $1.05$ is obtained for $31\,272$ degrees of freedom based on statistical uncertainties only,
indicating a relatively good quality of the fit~\cite{BABAR2010,Peng2014,BABAR2008,Bellesigmatwo,CDFsigmatwo}.

The time-dependent Dalitz plot analysis of \BtoDhzero with $D \to K_{S}^{0} \pi^{+} \pi^{-}$ decays is performed using data samples containing
$471 \times 10^{6}$ $\B\Bb$ pairs recorded with the \babar\ detector~\cite{BaBarDetector,BaBarluminosity} at the asymmetric-energy $e^+ e^-$ (3.1 on 9~GeV) collider PEP-II~\cite{PEPII} and
$772 \times 10^{6}$ $\B\Bb$ pairs recorded with the Belle detector~\cite{BelleDetector} at the asymmetric-energy $e^+ e^-$ (3.5 on 8~GeV) collider KEKB~\cite{KEKB} collected at the $\Upsilon(4S)$~\cite{MCgeneratorsandsimulations}.

The light neutral hadron $h^{0}$ is reconstructed in the decay modes $\piz \to \gamma \gamma$, $\eta \to \gamma \gamma$ and $\pip \pim \piz$, and $\omega \to \pip \pim \piz$.
Neutral \D mesons are reconstructed in the decay mode \DtoKSpipi, and neutral \Dstar mesons are reconstructed in the decay mode $\Dstar \to \D \piz$.
The decay modes $\Bz \to \D \piz$, $\D \eta$, $\D \omega$, $\Dstar \piz$, and $\Dstar \eta$, where sufficient signal yields are reconstructed, are included in the analysis.
The selection requirements applied to the reconstructed candidates are summarized in Ref.~\cite{accompanyingPRDpaper}.

\begin{figure*}[htb]
\includegraphics[width=0.95\textwidth]{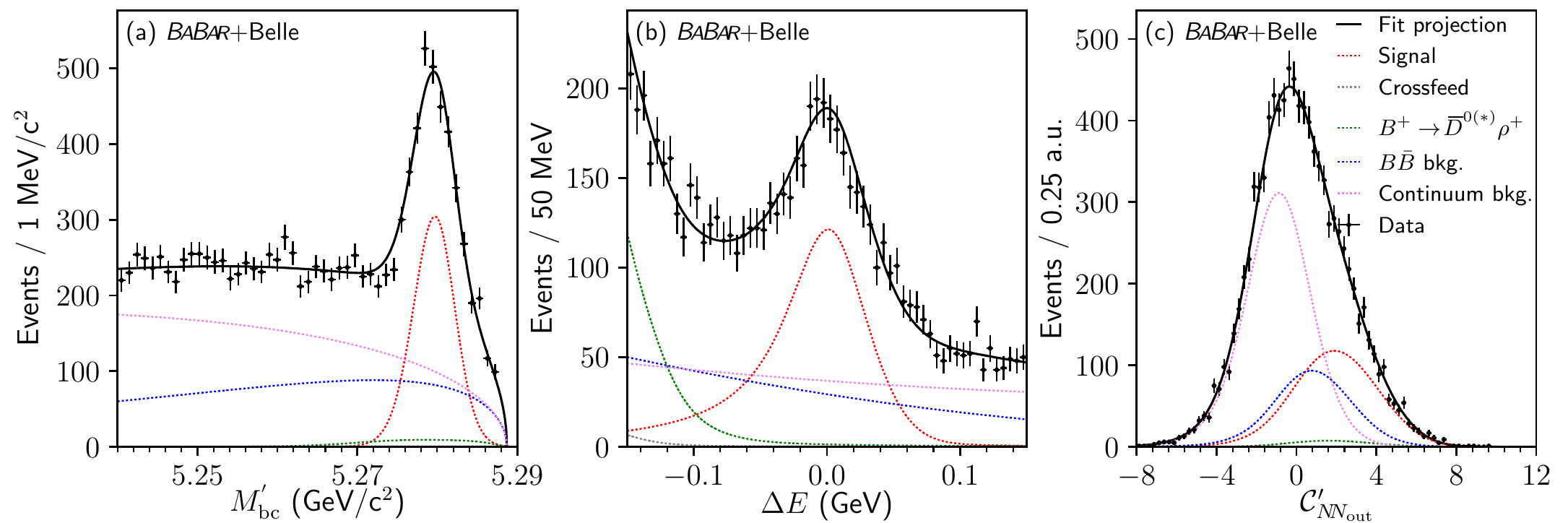}
\caption{
(color online). Data distributions for a) \Mbcprime, b) \DeltaE, and c) \NNoutprime (points with error bars) for the \babar\ and Belle data samples combined. The solid black lines represent projections of
the total fit function, and the colored dotted lines show the signal and background components of the fit as indicated in the legend.
In plotting the \Mbcprime, \DeltaE, and \NNoutprime distributions, each of the other two observables are required to satisfy
$\Mbcprime > 5.272\gevcc$, $\lvert\Delta E\rvert<100\mev$, or $0 < \NNoutprime < 8$ to select signal-enhanced regions.
}
\label{figure_Mbc_DeltaE_nbout_fit}
\end{figure*}

The \BtoDhzero yields are determined by three-dimensional unbinned maximum likelihood fits to the distributions of the observables \Mbcprime, \DeltaE, and \NNoutprime.
The beam-energy-constrained mass \Mbcprime defined in Ref.~\cite{Mbcprime_definition} is computed from the beam energy $E^{*}_{\rm beam}$ in the c.m.\ frame, the $D^{(*)}$ candidate momenta, and the $h^{0}$ candidate direction of flight.
The quantity \Mbcprime provides an observable that is insensitive to possible correlations with the energy difference $\Delta E = E^{*}_{\B} - E^{*}_{\rm beam}$ that can be induced by
energy mismeasurements for particles detected in the electromagnetic calorimeters, for example, caused by shower leakage effects.
The variable \NNoutprime defined in Ref.~\cite{NNoutprime_definition} is constructed from the output of a neural network multivariate classifier trained on event shape information based on a combination of 16 modified Fox-Wolfram moments~\cite{Neurobayes,FWmoments}
to identify background originating from $e^{+}e^{-} \to q \overline{q}$  $( q \in \{ u, d, s, c \} )$ continuum events.
The fit model accounts for contributions from \BtoDhzero signal decays, cross-feed from partially-reconstructed $\Bz \to \Dstar h^{0}$ decays,
background from partially-reconstructed $\Bp \to \overline{\kern -0.2em D}{}^{(*)0} \rho^{+}$ decays, combinatorial background from $\B\Bb$ decays, and background from continuum events.
In total, a \BtoDhzero signal yield of $ 1\,129 \pm 48 $ events in the \babar\ data sample and $ 1\,567 \pm 56 $ events in the Belle data sample is obtained.
The signal yields are summarized in Table IV of Ref.~\cite{accompanyingPRDpaper}.
The \Mbcprime, \DeltaE, and \NNoutprime data distributions and fit projections are shown in Fig.~\ref{figure_Mbc_DeltaE_nbout_fit}.

The time-dependent Dalitz plot analysis follows the technique established in the previous combined \babar+Belle time-dependent \CP violation measurement of $\Bzb \to D^{(*)}_{\CP} h^{0}$ decays~\cite{Roehrken2015}.
The measurement is performed by maximizing the log-likelihood function constructed from the events reconstructed from \babar\ and Belle data~\cite{accompanyingPRDpaper}.
The measurement includes all events used in the previous \Mbcprime, \DeltaE, and \NNoutprime fits.
In the log-likelihood function, the p.d.f.s are functions of the experimental flavor-tagged proper-time interval and Dalitz plot distributions for the signal and background components.
The signal p.d.f.s are constructed from Eqs.~\ref{equation:decay_rate} and~\ref{equation_sine_cosine} convolved with experiment-specific resolution functions to account for the finite vertex resolution~\cite{BaBar_btoccs,BelleVertexResolution}
and including the effect of incorrect flavor assignments~\cite{BaBar_btoccs,BelleTaggingNIM}.
The p.d.f.s for the proper-time interval distributions of the combinatorial background from $B\bar{B}$ decays and background from continuum events
account for background from non-prompt and prompt particles convolved with effective resolution functions.
The partially-reconstructed $\Bz \to \D^{*} h^{0}$ decays are modeled by the signal p.d.f. with a different set of parameters to account for this cross-feed contribution,
and the background from partially-reconstructed $\Bp \to \overline{\kern -0.2em D}{}^{(*)0} \rho^{+}$ decays is parameterized by an exponential p.d.f. convolved with the same resolution functions as used for the signal.

In the fit, the parameters $\tau_{\Bz}$, $\tau_{\Bp}$, and $\Delta m_{d}$ are fixed to the world averages~\cite{HFAG}, and the Dalitz plot amplitude model parameters are fixed to the results of the \DzerotoKSpipi Dalitz plot fit described above.
The signal and background fractions are evaluated on an event-by-event basis from the three-dimensional fit of the \Mbcprime, \DeltaE, and \NNoutprime observables.
The only free parameters are $\sin{2\beta}$ and $\cos{2\beta}$, and the results are
\begin{align}
\sin{2\beta} = 0.80 \pm 0.14  \,(\rm{stat.}) \pm 0.06 \,(\rm{syst.}) \pm 0.03 \,(\rm{model}) \mathrm{,} \nonumber \\
\cos{2\beta} = 0.91 \pm 0.22  \,(\rm{stat.}) \pm 0.09 \,(\rm{syst.}) \pm 0.07 \,(\rm{model}) \mathrm{.}
\end{align}
The second quoted uncertainty is the experimental systematic error, and the third is due to the \DzerotoKSpipi decay amplitude model.
The evaluation of these uncertainties is described in detail in Ref.~\cite{accompanyingPRDpaper}.
The linear correlation between $\sin{2\beta}$ and $\cos{2\beta}$ is $5.1\%$.
The result deviates less than $1.0$ standard deviations from the trigonometric constraint given by $\sin^{2}{2\beta} + \cos^{2}{2\beta} = 1$.

An alternative fit is performed to measure directly the angle $\beta$ using the signal p.d.f.\ constructed from Eq.~(\ref{equation:decay_rate}), and the result is
\begin{align}
\beta = \left( 22.5 \pm 4.4  \,(\rm{stat.}) \pm 1.2 \,(\rm{syst.}) \pm 0.6 \,(\rm{model}) \right)^{\circ} \mathrm{.}
\end{align}

\begin{figure}[htb]
\includegraphics[width=0.47\textwidth]{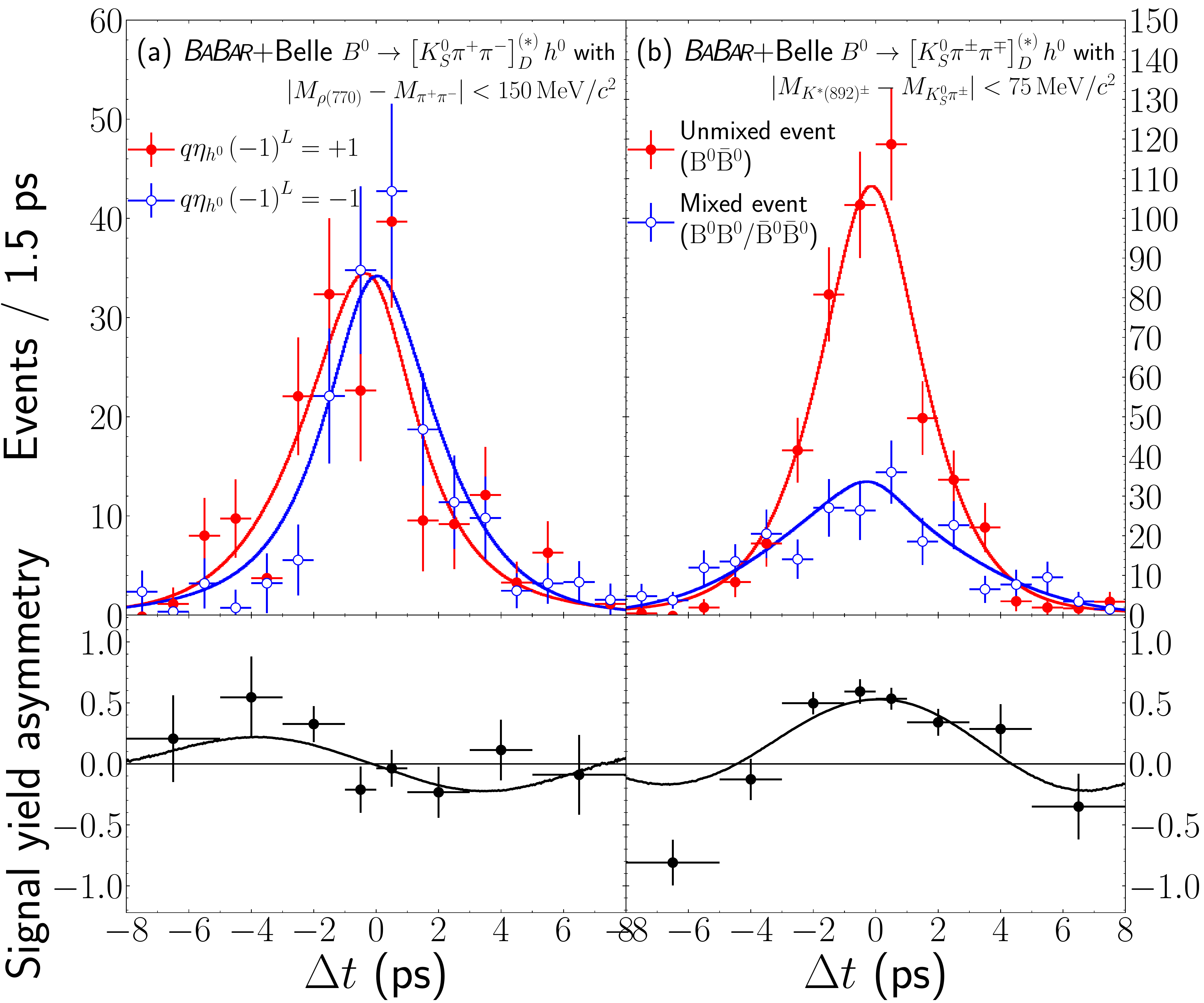}
\caption{
(color online). Distributions of the proper-time interval (data points with error bars) and the corresponding asymmetries for \BtoDhzero candidates associated
with high-quality flavor tags for two different regions of the \DtoKSpipi phase space and for the \babar\ and Belle data samples combined.
The background has been subtracted using the \sPlot technique~\cite{sPlotNIM}, with weights obtained from the fit presented in Fig.~\ref{figure_Mbc_DeltaE_nbout_fit}.
}
\label{figure:DeltaT_data_distributions_and_projections_of_the_fit}
\end{figure}

The proper-time interval distributions and projections of the fit for $\sin{2\beta}$ and $\cos{2\beta}$ are shown in Fig.~\ref{figure:DeltaT_data_distributions_and_projections_of_the_fit}
for two different regions of the \DzerotoKSpipi phase space.
Figure~\ref{figure:DeltaT_data_distributions_and_projections_of_the_fit}a shows a region predominantly populated by \CP eigenstates, $\Bz \to \left[ \KS \rho(770)^{0} \right]^{(*)}_{D} h^{0}$.
For these decays, interference emerges between the amplitude for direct decays of neutral \B mesons into these final states and those following \Bz-\Bzb oscillations.
The time evolution exhibits mixing-induced \CP violation governed by the \CP-violating weak phase $2\beta$, which manifests as a sinusoidal oscillation in the \CP asymmetry.
Figure~\ref{figure:DeltaT_data_distributions_and_projections_of_the_fit}b shows a region predominantly populated by quasi-flavor-specific decays, $\Bz \to \left[ K^{*}(892)^{\pm} \pi^{\mp} \right]^{(*)}_{D} h^{0}$.
For these decays, the time evolution exhibits \Bz-\Bzb oscillations governed by the oscillation frequency, $\Delta m_{d}$,
which appears as an oscillation proportional to $\cos(\Delta m_{d} \Delta t)$ in the corresponding asymmetry.

The measurement procedure is validated by various cross-checks.
The $\Bz \to \Db^{(*)0} h^{0}$ decays with the CKM-favored $\Dzb \to \Kp \pim$ decay have very similar kinematics and background composition as \BtoDhzero with \DtoKSpipi decays and provide a high-statistics control sample.
Using the same analysis approach, the time-dependent \CP violation measurement of the control sample results in mixing-induced and direct \CP violation consistent with zero, in agreement with the assumption of negligible \CP violation for these flavor-specific decays. 
Measurements of the neutral \B meson lifetime for \BtoDhzero with \DtoKSpipi decays, and for the control sample without flavor-tagging applied, yield
$\tau_{\Bz} = \left(1.500 \pm 0.052\,(\rm{stat.})\right)\,\mathrm{ps}$ and $\tau_{\Bz} = \left(1.535 \pm 0.028\,(\rm{stat.})\right)\,\mathrm{ps}$, respectively, which are in agreement with the world average $\tau_{\Bz} = \left(1.520 \pm 0.004\right)\,\mathrm{ps}$~\cite{HFAG}.
In addition, we have performed all measurements for data separated by experiment yielding consistent results~\cite{accompanyingPRDpaper}.

The significance of the results is determined by a likelihood-ratio approach that accounts for the experimental systematic uncertainties and the Dalitz plot amplitude model uncertainties
by convolution of the likelihood curves.
The measurement of $\sin{2\beta}$ agrees within $0.7$ standard deviations with the world average of $\sin{2\beta} = 0.691 \pm 0.017$~\cite{HFAG} obtained from more precise measurements using $\bar{b} \to \bar{c}c\bar{s}$ transitions.
The measurement of $\cos{2\beta}$ excludes the hypothesis of $\cos{2\beta}\leq 0$ at a $p$-value of $2.5 \times 10^{-4}$,
which corresponds to a significance of $3.7$ standard deviations, providing the first evidence for $\cos{2\beta}>0$.
The measurement of $\beta$ excludes the hypothesis of $\beta=0^{\circ}$ at a $p$-value of $3.6 \times 10^{-7}$, which corresponds to a significance of $5.1$ standard deviations.
Hence, we report an observation of \CP violation in \BtoDhzero decays.
The result for $\beta$ agrees well with the preferred solution of the Unitarity Triangle, which is $(21.9 \pm 0.7)^{\circ}$, if computed from the world average of $\sin{2\beta} = 0.691 \pm 0.017$~\cite{HFAG}.
The measurement excludes the second solution of $\pi/2 - \beta = (68.1 \pm 0.7)^{\circ}$ at a $p$-value of $2.31 \times 10^{-13}$, corresponding to a significance of $7.3$ standard deviations.
Therefore, the present measurement resolves an ambiguity in the determination of the apex of the CKM Unitarity Triangle.

In summary, we combine the final \babar\ and Belle data samples, totaling an integrated luminosity of more than $1\,\mathrm{ab}^{-1}$ collected at the $\Upsilon\left(4S\right)$ resonance,
and perform a time-dependent Dalitz plot analysis of \BtoDhzero with \DtoKSpipi decays.
We report the world's most precise measurement of the cosine of the \CP-violating weak phase $2 \beta$ and obtain the first evidence for $\cos{2\beta}>0$.
The measurement directly excludes the trigonometric multifold solution of $\pi/2 - \beta = (68.1 \pm 0.7)^{\circ}$ without any assumptions, and thus resolves an ambiguity related to the CKM Unitarity Triangle parameters.
An observation of \CP violation in \BtoDhzero decays is reported.

The \BtoDhzero decays studied by the combined \babar\ and Belle approach provide a probe for the \CP-violating weak phase $2\beta$ that is
theoretically more clean than the ``gold plated'' decay modes mediated by $\bar{b} \to \bar{c}c\bar{s}$ transitions~\cite{Fleischer2003}.
Therefore, \BtoDhzero decays can provide a new and complementary SM reference for $2\beta$ at the experimental precision
achievable by the future high-luminosity \B factory experiment Belle II~\cite{BelleIITDR}.

We thank the \pep2\ and KEKB groups for the excellent operation of the accelerators.
The \babar\ experiment acknowledges the substantial dedicated effort from the computing organizations for their support.
The collaborating institutions wish to thank SLAC for its support and kind hospitality. 
The Belle experiment wishes to acknowledge the KEK cryogenics group for efficient solenoid
operations; and the KEK computer group, the NII, and 
PNNL/EMSL for valuable computing and SINET5 network support.
This work was supported by
MEXT, JSPS and Nagoya's TLPRC (Japan);
ARC (Australia);
FWF (Austria);
NSERC (Canada);
NSFC and CCEPP (China); 
MSMT (Czechia);
CEA and CNRS-IN2P3 (France);
BMBF, CZF, DFG, EXC153, and VS (Germany);
DST (India); INFN (Italy); 
MOE, MSIP, NRF, RSRI, FLRFAS project and GSDC of KISTI (Korea);
FOM (The Netherlands);
NFR (Norway);
MNiSW and NCN (Poland);
MES and RFAAE (Russia);
ARRS (Slovenia);
IKERBASQUE and MINECO (Spain); 
SNSF (Switzerland);
MOE and MOST (Taiwan);
STFC (United Kingdom);
BSF (USA-Israel);
and DOE and NSF (USA).
Individuals have received support from the
Marie Curie EIF (European Union)
and the A.~P.~Sloan Foundation (USA).


\begin{thebibliography}{99}
\bibitem{phi_three} The CKM angle $\gamma$ is also referred to as $\phi_3$ in the literature.
\end{thebibliography}
\end{document}